\documentclass[fleqn,usenatbib]{mnras}
\usepackage{newtxtext,newtxmath}
\usepackage[T1]{fontenc}
\DeclareRobustCommand{\VAN}[3]{#2}
\let\VANthebibliography\thebibliography
\def\thebibliography{\DeclareRobustCommand{\VAN}[3]{##3}\VANthebibliography}
\usepackage{graphicx}	
\usepackage{amsmath}	
\usepackage{multirow}
\usepackage[normalem]{ulem}

\title[Short-Period Decayless Kink Oscillations]{Two-Spacecraft Detection of Short-period Decayless Kink Oscillations of Solar Coronal Loops}

\author[S. Zhong et al.]{
Sihui Zhong,$^{1}$
Valery M. Nakariakov,$^{1,2}$\thanks{E-mail: v.nakariakov@warwick.ac.uk}
Dmitrii Y. Kolotkov,$^{1}$
Cis Verbeeck,$^{3}$
David Berghmans$^{3}$
\\
$^{1}$ Centre for Fusion, Space and Astrophysics, Department of Physics, University of Warwick, Coventry CV4 7AL, UK\\
$^{2}$ Centro de Investigacion en Astronom\'ia, Universidad Bernardo O'Higgins, Avenida Viel 1497, Santiago, Chile\\
$^{3}$ Solar-Terrestrial Centre of Excellence - SIDC, Royal Observatory of Belgium, Ringlaan -3- Av. Circulair, 1180 Brussels, Belgium\\}

\date{Accepted XXX. Received YYY; in original form ZZZ}

\pubyear{2022}

\begin{document}
\label{firstpage}
\pagerange{\pageref{firstpage}--\pageref{lastpage}}
\maketitle

\begin{abstract}
Decayless kink oscillations of an ensemble of loops are captured simultaneously by the High Resolution Imager (HRI) of the Extreme Ultraviolet Imager (EUI) and the Atmospheric Imaging Assembly (AIA) from 22:58 UT on 5 November to 00:27 UT on 6 November 2021.
Oscillations are analysed by processing image sequences taken by the two instruments with a motion magnification technique.
The analysed loops are around 51\,Mm in length, and oscillate with short periods of 1--3\,min (1.6\,min in average) and displacement amplitudes of 27--83\,km. 
The signals recorded by AIA are delayed by 66\,s as compared to HRI, which coincides with the light travel time difference from the Sun to each instrument. After correction of this time difference, the cross-correlation coefficient between the signals from the two data varies from 0.82 to 0.97, indicating that they are well consistent.
This work confirms that HRI sees the same oscillations as AIA, which is the necessary first step before proceeding to the detection of shorter time scales by EUI. In addition, our results indicate the robustness of the de-jittering procedure in the study of kink oscillations with HRI.

\end{abstract}




\section{Introduction}
\label{sec:intro}

Oscillatory and wave phenomena detected in the solar corona are subject to intensive study mainly in the context of their possible relevance to the long-standing problem of coronal plasma heating \citep[e.g.][]{2020SSRv..216..140V} and because of their plasma diagnostics potential \citep[e.g.][]{2021SSRv..217...73N}. A puzzling class of coronal oscillations are the omnipresent low-amplitude decayless kink oscillations of coronal loops, appearing as repetitive displacements of the loops in the plane of the sky \citep{2012ApJ...751L..27W} or as persistent Doppler shift oscillations of coronal emission lines \citep{2012ApJ...759..144T}. The oscillation periods have been found to range from 1.5 to 10~min, and scale linearly with the loop length \citep{2015A&A...583A.136A}. The latter finding indicates that the oscillations are standing kink modes. The average apparent displacement amplitude in the plane of the sky is 0.17~Mm, i.e., smaller than the pixel size of available coronal EUV imagers.

The main interest in decayless kink oscillations is connected with their ubiquity. In particular, the oscillations are seen to occur without any association with energy releases such as flows or eruptions. It allows one to use them for seismological diagnostics of coronal loops and active regions during quiet times 
\citep{2019ApJ...884L..40A}, i.e., before flares and eruptions. The simultaneous detection of the fundamental and second parallel (axial) harmonics \citep{2018ApJ...854L...5D} opened up a possibility to study the field-aligned structure of the equilibrium plasma parameters in the oscillating loop \citep[e.g.][]{2005ApJ...624L..57A,2009SSRv..149....3A}. In addition, revealing the mechanism for the compensation of the oscillation damping may shed light on the energy balance in active regions.

Several mechanisms responsible for the existence of decayless kink oscillations have been proposed. \citet{2016A&A...591L...5N} ruled out their excitation by monochromatic drivers such as p-modes or chromospheric oscillations. Concerning non-monochromatic mechanisms, it has been proposed that the energy could come from slowly-varying flows by the self-oscillatory mechanism \citep{2016A&A...591L...5N, 2020ApJ...897L..35K}, or the Alfv\'enic vortex shedding \citep{2009A&A...502..661N, 2021ApJ...908L...7K}, or random motions around footpoints of the oscillating loop \citep{2020A&A...633L...8A, 2021MNRAS.501.3017R, 2021SoPh..296..124R}. Apparently decayless patterns could also appear because of the development of the Kelvin--Helmholtz instability \citep[e.g.,][]{2016ApJ...830L..22A, 2019ApJ...870...55G}.

The recent commissioning of the High Resolution Imager (HRI) of the Extreme Ultraviolet Imager (EUI, \citealt{2020A&A...642A...8R}) onboard Solar Orbiter \citep[][]{2020A&A...642A...1M}, opens up interesting perspectives in the high-resolution study of the decayless kink oscillation phenomenon. This extends beyond the resolution of the Atmospheric Imaging Assembly (AIA, \citealt{2012SoPh..275...17L}) on the Solar Dynamics Observatory (SDO, \citealt{2012SoPh..275....3P}), which is currently used for time--distance analysis of these oscillations.
In particular, one may anticipate to resolve higher parallel harmonics which should result naturally from a broadband driver \citep[e.g.][]{2018ApJ...854L...5D, 2020A&A...633L...8A}, and obtain more robust information about the variability of instantaneous parameters of the oscillation \citep[e.g.][]{2022MNRAS.513.1834Z} and the oscillating loops \citep[e.g.][]{2019FrASS...6...38K, 2021ApJ...922...60S}.

In this paper, we present the simultaneous detection of decayless kink oscillations with HRI and AIA, which has been successfully used for the observations of decayless kink oscillations for almost ten years. The oscillation periods detected in this study vary from 1-3\,min, which is in the short limit of the previous detections by AIA. We consider this study as the necessary step towards the follow-up progression to shorter time scales detected by HRI, which are not possible with AIA.

\section{Observations and Data Analysis}
\label{sec:Obs}

\begin{figure}
	\includegraphics[width=\columnwidth]{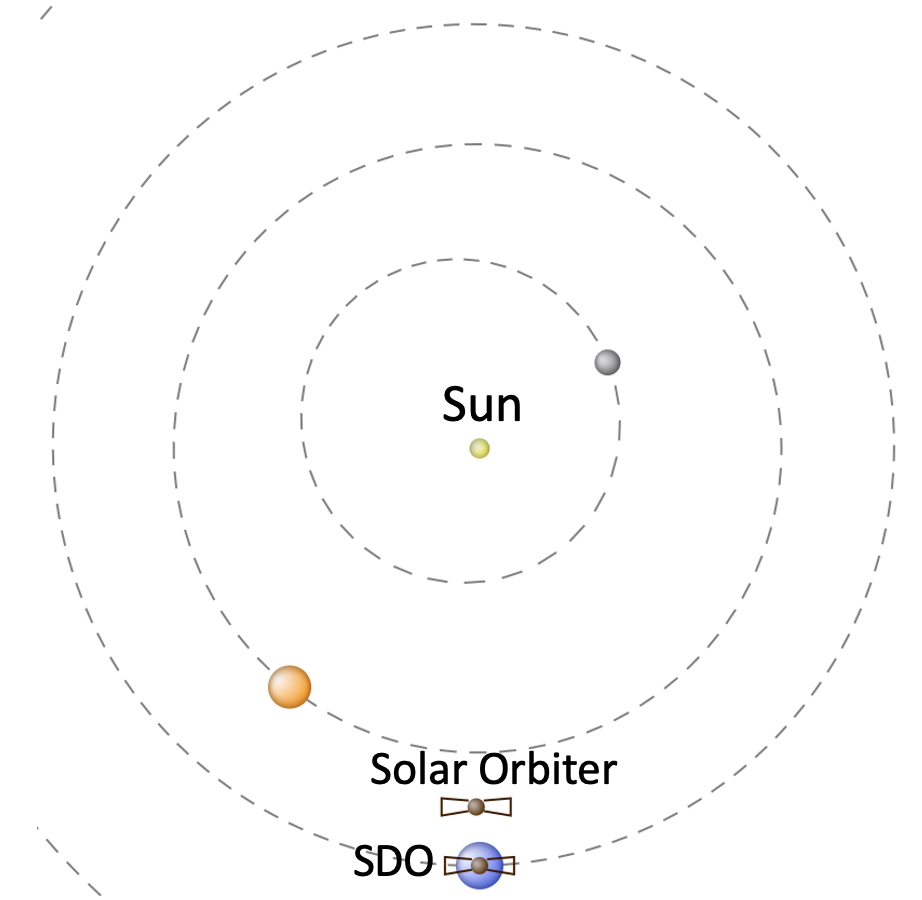}
    \caption{The locations of Solar Orbiter and SDO at 22:58 UT on 2021 November 5, seen in the ecliptic plane. The grey, orange, and blue circles with dashed orbit are Mercury, Venus, and Earth, respectively. This is a screenshot from the propagation tool (\url{http://propagationtool.cdpp.eu/}).
	}
    \label{fig:los}
\end{figure}

A series of 174 \AA\, images was obtained with HRI from 22:58~UT on 5 November 2021 to 00:28~UT on
6 November 2021. The field of view (FOV) was 2048$\times$2048 pixels with the pixel size of 0.492\arcsec\ and the time cadence of 5\,s. In this study, we use level 2 data\footnote{The EUI L2 data we used can be accessed via \lq\lq EUI Data Release 5.0 2022-04\rq\rq, see \url{https://doi.org/10.24414/2qfw-tr95}.}. 
The HRI telescope was pointed at near the disk centre, where a bundle of loops with ongoing decayless kink oscillations was situated at an active region. 
Simultaneous observation was acquired by SDO/AIA at 171\AA\ with a 
pixel size of 0.6\arcsec\ and time cadence of 12\,s. 
At the time of observation, the Solar Orbiter was at a distance of 0.86\,au from the Sun, at the Stonyhurst heliographic longitude of -0.574\degr, and the latitude of 1.987\degr; while SDO was at a distance of 0.99\,au, longitude of -0.0129\degr, latitude of 3.875\degr.
In this case, see Fig.\ref{fig:los}, both spacecraft have nearly parallel lines of sight (LoS). 
Also, such a location determines the linear plate scale of HRI images as 306\,km pixel$^{-1}$, while that of AIA images is 435\,km pixel$^{-1}$.
With higher spatial resolution and double temporal resolution, HRI allows us to detect short-period low-amplitude kink oscillations of plasma structures, but our aim is to study the oscillation detected simultaneously by both HRI and AIA.
We select a similar 171\,\AA\ wavelength in AIA in comparison with HRI 174\,\AA\ to confirm the oscillation detection. 
In addition, magnetograms from the Helioseismic and Magnetic Imager (HMI, \citealt{2012SoPh..275..207S}), with resolution of 0.5\arcsec\ pixel$^{-1}$ and cadence of 45~s, are used to establish the magnetic connectivity of the loops of interest.

\subsection{Alignment}
\label{sec:align1}

After compensating the solar rotation in all data sets, it is found that the pointing accuracy of HRI images
as defined in the image metadata (FITS keywords)
is not sufficient for the present study.
So the HRI images need to be
internally aligned to remove spacecraft jitter. 

One frequently used method to do image alignment is the cross-correlation technique \citep[][]{ZITOVA2003977}. 
This method first calculates the cross-power spectrum of the reference image and aligned image, and returns a cross correlation surface/matrix of the same size as the image through the inverse Fourier transform. The distance from the location of the maximum correlation to the surface centre in Cartesian axes is the offset in the $x$ and $y$ directions.
Given that the correlation matrix is in units of pixels, the location of the maximum value is in a certain pixel. To achieve the sub-pixel accuracy, we interpolate the correlation peak \citep[][]{Lehmann1999}.
The corresponding SSWIDL function is \texttt{tr\_get\_disp.pro}, 
which is the core block for the later developed routines  \texttt{align\_cube\_correl.pro} and \texttt{fg\_rigidalign.pro}. 
The accuracy of \texttt{fg\_rigidalign.pro} is several tenths of a pixel, see \url{https://hesperia.gsfc.nasa.gov/ssw/trace/idl/util/tr_get_disp.pro}. 
Alternatively, upsampling the correlation surface can also result in sub-pixel accuracy \citep[e.g.,][]{Guizar-Sicairos:2008,2017Jaime}.

Recently, \citet{6463241} developed an algorithm to locate the correlation peak with high accuracy by treating the centroid of the correlation surface as the peak. 
This algorithm first calculates the displacement of the pixel-level and shifts the aligned image, then it performs iteration of the offset calculation and shifting until the displacement is less than one pixel, finally locates the centroid with accuracy as high as 0.01 pixel (i.e., 36\,km in the analysed data, i.e., $1/5$ of average amplitude of decayless oscillations 170\,km) and shifts the sub-pixel displacement \citep[][]{2015RAA....15..569Y}.
Considering its high accuracy, we apply this method to align the HRI data in this study.

\subsection{Co-alignment and motion magnification}
\label{sec:coalignment}
After the internal alignment, co-alignment between HRI images and AIA images closest in time is performed based on the common features in both data sets. HMI data is aligned with AIA already. 
For further analysis, sub-frames are extracted at the same FOV (indicated by the red box in Fig.\ref{fig:FOV}(b)) and stacked into a 3D datacube, respectively.

At 22:58 UT on 5 November 2021, the HRI telescope captures a scene near the disk centre, including 
active region AR 12983 with several bundles of loops evolving. During this 1.5-hour observation, decayless kink oscillations of an ensemble of loops are observed, and no flares or CMEs are reported in this region.
A similar scene is simultaneously recorded by AIA. 
As the characteristic displacement amplitude (0.17\,Mm) of decayless kink oscillations is less than the pixel size of both datasets, we use the motion magnification technique to magnify the transverse motions in the plane of the sky \citep[][]{2016SoPh..291.3251A,2021SoPh..296..135Z}. In this work, the magnification factor is 3, and smoothing width is 4 minutes, which is longer than the oscillation periods of interest.
Having applied the motion magnification to HRI and AIA datasets, we make time--distance (TD) plots (Fig.~\ref{fig:td}) using slits (see white slits in Fig.~\ref{fig:evo}) directed across the oscillating loops, to reveal the oscillatory patterns \citep[see][ for the methodological details]{2022MNRAS.513.1834Z}. 

\section{Results}
\label{sec:result}

\subsection{Oscillation detection}
\label{sec:co}

\begin{figure*}
	\includegraphics[width=\textwidth]{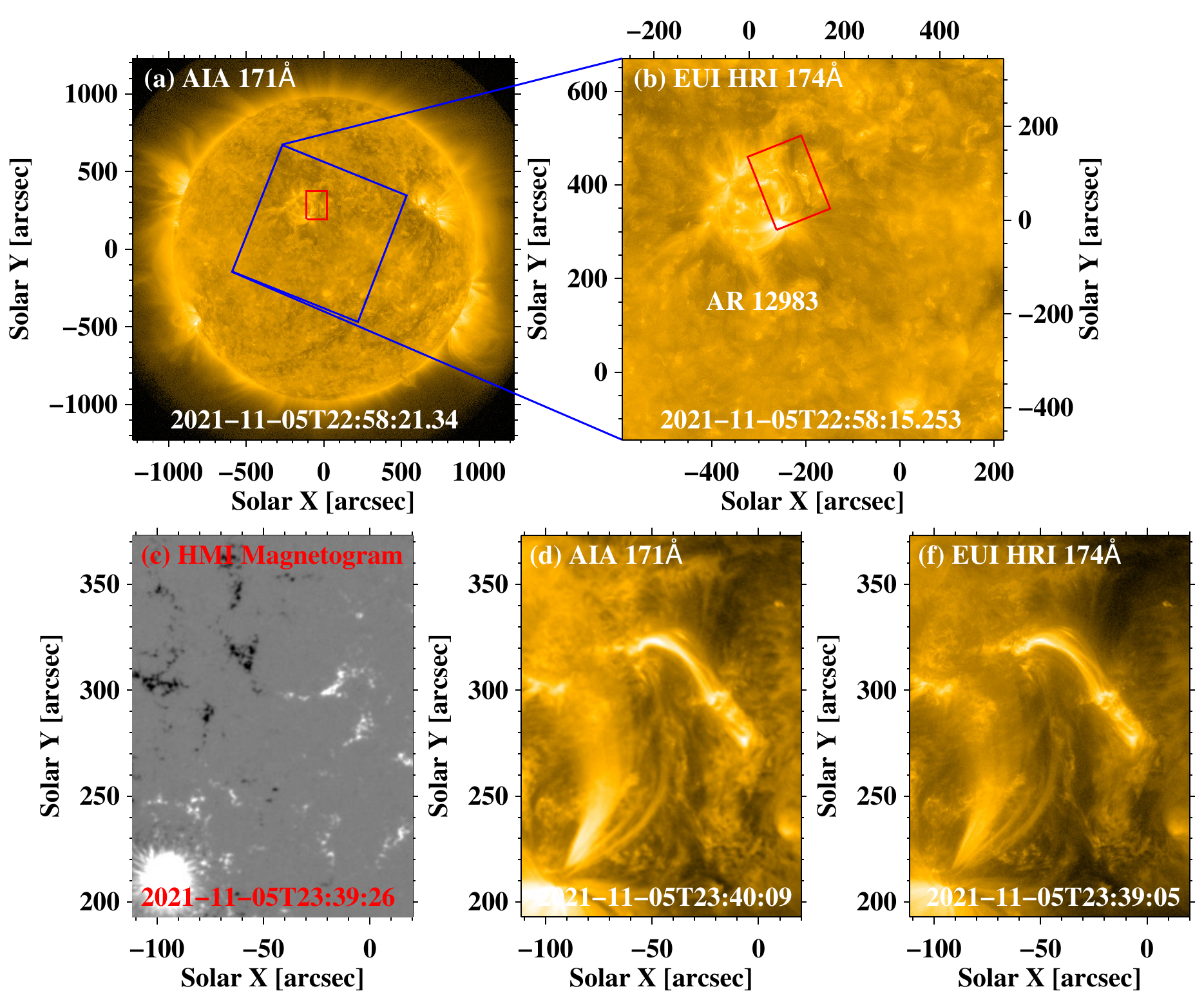}
    \caption{Panel (a): Full-disk image in the AIA 171\,\AA\, passband. Panel (b): HRI 174\,\AA\,image, whose field of view (FOV) is overplotted on panel (a) as the blue box. 
    Bottom panels: HMI magetogram, AIA 171\,\AA\,image, and an HRI image extracted in the FOV of interest. All three images are aligned.
    The FOV of HRI images are overplotted on the full-disk  171\,\AA\,image. 
    The red rectangles in panels (a)-(b) indicate the FOV of interest.
	}
    \label{fig:FOV}
\end{figure*}

Fig.~\ref{fig:FOV} shows the full-disk AIA 171\,\AA\, image (panel (a)) and HRI 174
\,\AA\, image (panel (b)) obtained at the beginning of the observation. The FOV of HRI is indicated by the blue box, and the red rectangles indicate the FOV of panels (c)-(e), which is the region of interest where the  oscillations occur. 
The bottom panels of Fig.~\ref{fig:FOV} display the HMI magnetogram (c), AIA 171\,\AA image (d), and HRI 174\,\AA image (e), all of which are aligned. 
The sunspot is situated at the bottom-left corner of FOV.  
The oscillations occur in the bright loops near [-50\arcsec, 320\arcsec], connecting two small magnetic patches of opposite polarities.
The apparent distance between the loop footpoints is 32\,Mm, hence the loop length is estimated as $L=\pi \times 32/2 =51\pm$16\,Mm, assuming that the loop shape is semi-circular. See also the online animation which shows the co-aligned HRI 174 \,\AA\, images (left) and AIA 171 \,\AA\, images (right) with time corrected. 

\begin{figure*}
	\includegraphics[width=\textwidth]{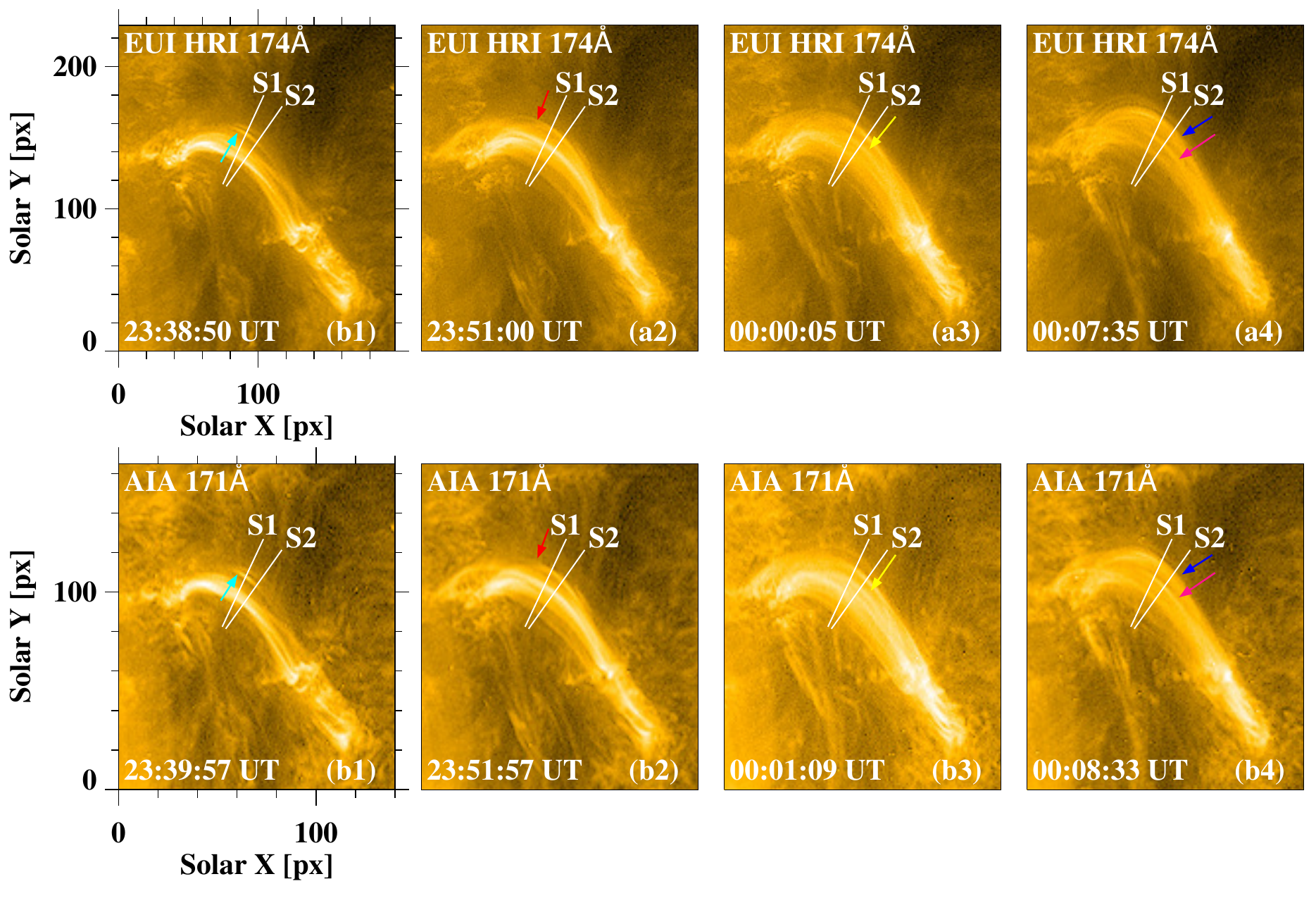}
    \caption{The dynamics of analysed threads. Slits \lq\lq S1\rq\rq\ and \lq\lq S2\rq\rq\ are used to make time--distance (TD) maps in Fig.~\ref{fig:td}. The coloured arrows denote the oscillating threads which have distinct oscillatory patterns. The images are enhanced to reveal the threads, by applying a high-boost mask.}
    \label{fig:evo}
\end{figure*}

\begin{figure*}
	\includegraphics[width=\textwidth]{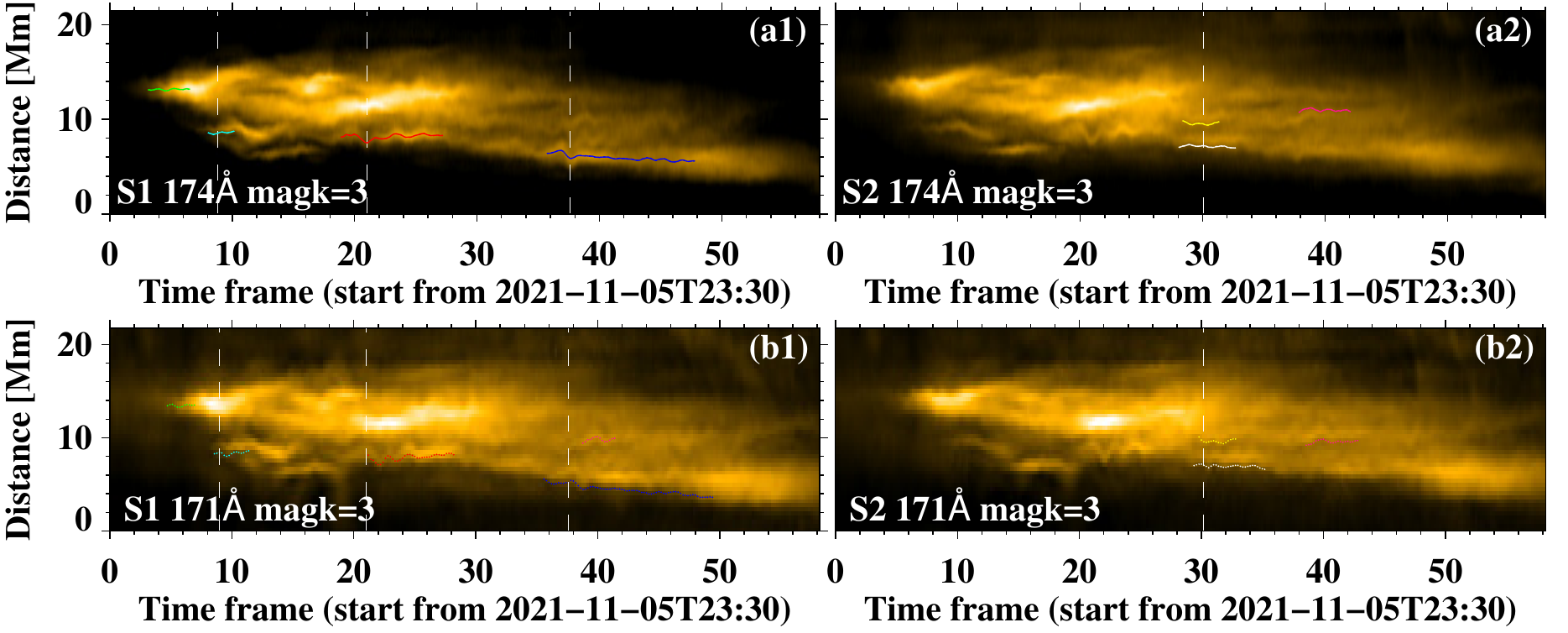}
    \caption{TD maps which display decayless kink oscillations. 
    The dashed lines denote the observation time of HRI images in Fig.~\ref{fig:evo}.
    The coloured curves indicate the displacement location of centre/edge of corresponding oscillating threads in Fig.~\ref{fig:evo}.
    The magnification factor is 3 for both data sets.}
    \label{fig:td}
\end{figure*}

The bundle of loops which hosts oscillations is shown in Fig.~\ref{fig:evo}. The images of upper and bottom rows are made with HRI and AIA respectively. 
Besides the obvious difference in image quality and despite the slightly different bandpasses, the images appear very similar.
Within this bundle of loops, several distinguishable threads (denoted by the coloured arrows) evolve quickly, resulting in that the observed kink oscillations appear to be short-lived. Two slits \lq\lq S1\rq\rq\ and \lq\lq S2\rq\rq\  across the loops are used to make TD plots to reveal the oscillatory patterns. As shown in Fig.~\ref{fig:td}, all oscillations last for only several cycles, up to 15 minutes. 
The disappearance of oscillations is caused by changing of observational conditions, e.g., other loops come into the LoS, rather than damping.
The oscillation signal of a certain thread is highlighted by curves of the same colour. These curves, marking the location of the loop centre or boundary at each instant of time, are determined by fitting the transverse loop profile for each instance of time with a Gaussian profile by \texttt{gaussfit.pro}.

To estimate the oscillation parameters, each oscillating signal is best fitted by a harmonic function with a linear background trend:
\begin{equation} 
    A(t)=A_{0}\sin \left(\frac{2\pi t}{P}+\phi \right)+c_{0}+c_{1}t.
	\label{eq:sine}
\end{equation}
where $A_0$ is the displacement amplitude, $P$ is the oscillation period, $\phi$ is the phase, and $c_{0}$ and $c_{1}$ are constants. 
The fitting results are summarised in Table~\ref{tab:pars}. 
The details of best fitting curves are given in Appendix~\ref{appendix}.
Note that here the oscillation amplitude is linearly magnified by factor 3, so the original amplitude is $A = A_0/3$. For convenience, the oscillating threads are numbered in the chronological order, i.e., from Thread 1 (T1 for short) to T7, corresponds to the green, cyan, red, yellow, white, blue, pink threads.
The oscillation periods of T1 to T7 are within 1-3 minutes, ranging from 67 to 133\,s. 
Compared with the previous observed events \citep[e.g.,][]{2015A&A...583A.136A,2022MNRAS.513.1834Z}, the oscillation periods in this study are rather short (see Section.~\ref{sec:period}). 
In addition, displacement amplitudes vary from 27\,km to 83\,km, which is much lower than the average amplitude of 170\,km \citep{2015A&A...583A.136A}.
Note that the values of the parameters obtained in HRI and AIA are consistent, within error bars. 
The averaged periods from T1 to T7 are $84\pm21$\,s, $74\pm9$\,s, $104\pm11$\,s, $83\pm16$\,s, $87\pm 9$\,s, $111\pm33$\,s, and $108\pm6$\,s, respectively.
The averaged amplitudes from T1 to T7 are $34\pm14$\,km, $41\pm13$\,km, $71\pm14$\,km, $59\pm 18$\,km, $35\pm9$\,km, $63\pm16$\,km, $41\pm14$\,km.


\subsection{Correlation of oscillations in two data}
\label{sec:correlation}

\begin{figure}
	\includegraphics[width=\columnwidth]{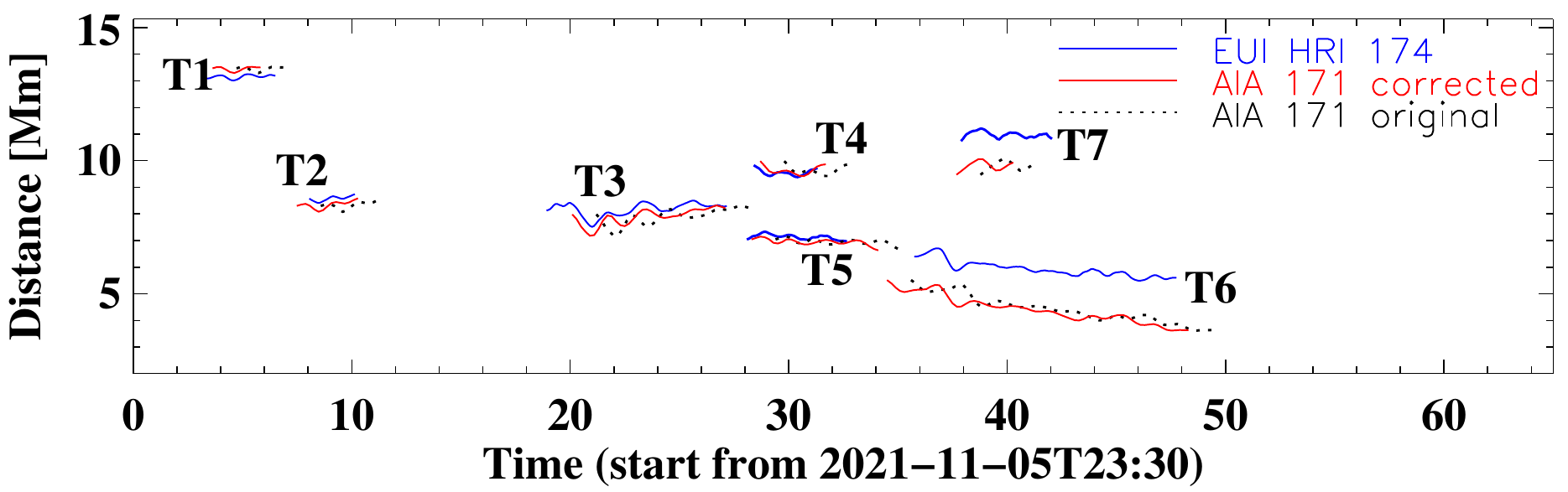}
    \caption{All oscillation signals obtained from HRI 174\,\AA\, images (blue) and AIA 171\,\AA\, images (red and black). The solid red curves are corrected signals with light travel time difference compensation. The black dotted curves are original uncorrected signals.}
    \label{fig:sum}
\end{figure}

Signals from HRI (see Fig.~\ref{fig:td}(a1)-(a2)) and AIA (see Fig.~\ref{fig:td}(b1)-(b2)) are very similar in shape but AIA signals are delayed by around 1 minute. 
During the observation, the distance difference of two spacecrafts from the Sun was $1.98\times 10^{7}$\,km, giving the light travel time of 66\,s. That is to say, the EUI received light 66\,s earlier than AIA. Taken this into account, we correct the AIA signals by shifting the signal forward by 66\,s. As shown in Fig.~\ref{fig:sum}, the corrected AIA and HRI signals are consistent with each other, both in space and time. 
To quantify the correlation of the oscillations in two data, the cross-correlation coefficient between the HRI and corrected AIA signal is calculated for each thread and displayed in  Table~\ref{tab:pars}.
The cross-correlations coefficient between signals from two data sets for T1 to T7 is 0.94, 0.97, 0.92, 0.96, 0.82, 0.95, 0.97, respectively. This indicates that oscillations seen with the two instruments correlate well with each other.

\begin{table}
	\centering
	\caption{Oscillation parameters of different oscillating thread estimated in HRI and AIA data. Information includes the thread ID, data source, period ($P$ in seconds), displacement amplitude ($A$ in km, the magnified amplitude divided by magnification factor, i.e., $A=A_{0}/3$), and cross-correlation coefficient (C) of signals from these two data sets.}
	\label{tab:pars}
	\begin{tabular}{lcccr} 
		\hline
		Thread & Data & $P$ [s] & $A$ [km] & C\\
		\hline
		\multirow{2}{*}{T1} & HRI 174 & $76\pm11$ & $24\pm7$ & \multirow{2}{*}{0.94}\\
		                   & AIA 171 & $91\pm30$ & $45\pm 22$ &\\
		 \hline
		\multirow{2}{*}{T2} & HRI 174 & $67\pm8$ & $28\pm 5$ & \multirow{2}{*}{0.97}\\
	                     	& AIA 171 & $80\pm10$ & $53\pm 22$ & \\
		\hline
		\multirow{2}{*}{T3}  & HRI 174 & $107\pm8$ & $70\pm9$ & \multirow{2}{*}{0.92}\\
		                     & AIA 171 & $100\pm13$ & $72\pm19$ & \\
		\hline
		\multirow{2}{*}{T4}  & HRI 174 & $80\pm8$ & $47\pm6$ & \multirow{2}{*}{0.96}\\
		                     &  171 & $86\pm 24$ & $71\pm30$ & \\
		\hline
		\multirow{2}{*}{T5}  & HRI 174 & $83\pm 9$ & $28\pm5$ & \multirow{2}{*}{0.82}\\
		                     & AIA 171 & $90\pm 8$ & $42\pm 13$ & \\
		\hline
		\multirow{2}{*}{T6}  & HRI 174 & $104\pm5$ & $31\pm10$ & \multirow{2}{*}{0.95}\\
		                     & AIA 171 & $112\pm6$ & $51\pm17$ & \\
		\hline
		 \multirow{2}{*}{T7}  & HRI 174 & $89\pm10$ & $44\pm5$ & \multirow{2}{*}{0.97}\\
		                      & AIA 171 & $133\pm55$ & $83\pm26$ & \\
		\hline
	\end{tabular}
\end{table}

\subsection{Short periodicity}
\label{sec:period}

\begin{figure}
	\includegraphics[width=\columnwidth]{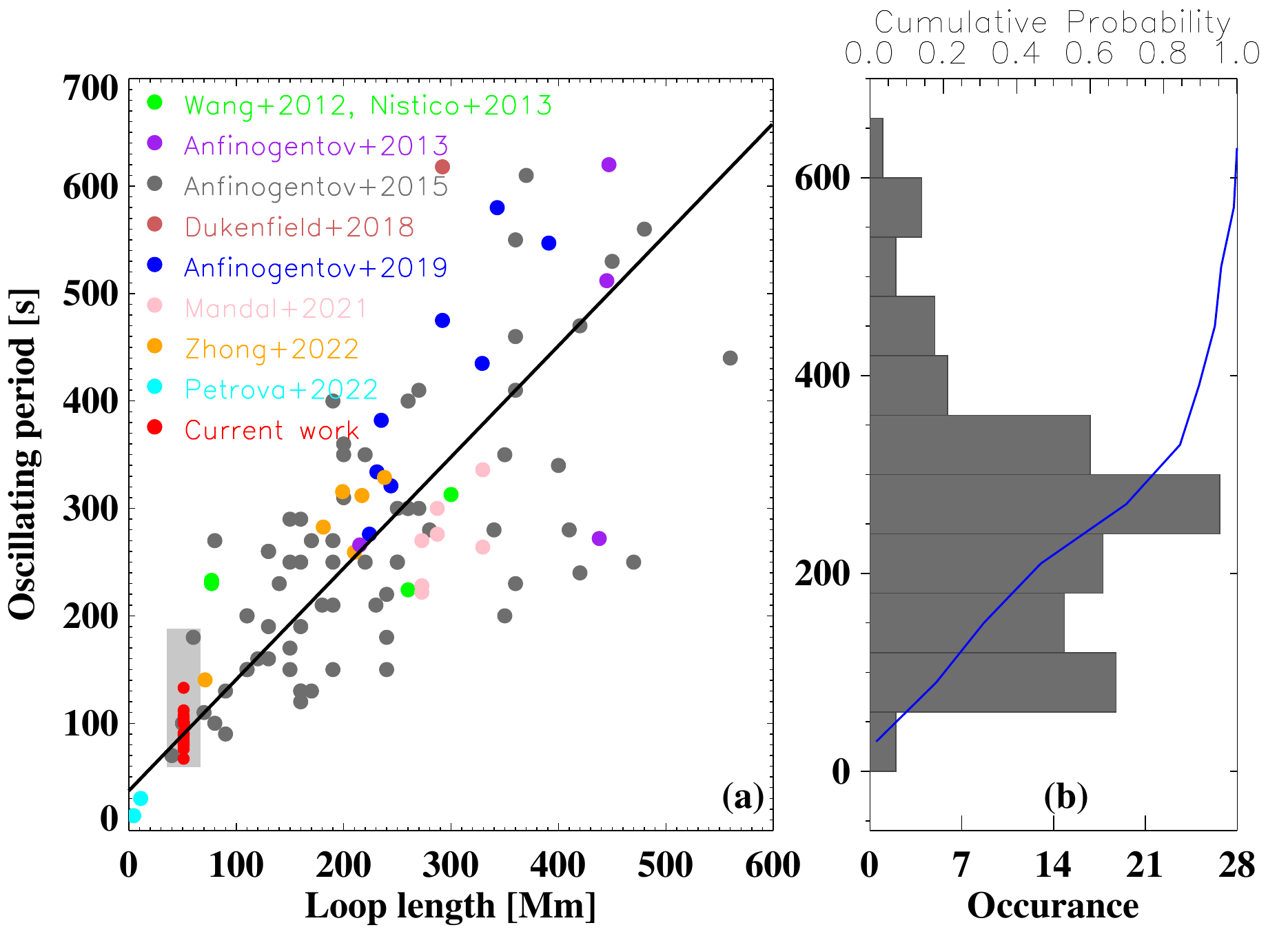}
    \caption{Scaling of the oscillation period and loop length  (a) and histogram of periods (b) based on the current work and previous studies on decayless kink oscillations of coronal loops.  
    Data-pairs in red are obtained in the current study. The grey region in shade represents the range of parameters in the current work. Data-pair shown in green, purple, grey, indian red, blue, pink, orange, and cyan, are collected from \citet{2012ApJ...751L..27W,2013A&A...552A..57N,2013A&A...560A.107A, 2015A&A...583A.136A,2018ApJ...854L...5D,2019ApJ...884L..40A,2021A&A...652L...3M,2022MNRAS.513.1834Z,2022arXiv220505319P}, respectively. 
    The black line is the best linear fit with the gradient of about unity. The bin size of histogram is 1\,min. The blue curve is the cumulative probability. Noting that the loop lengths of the oscillating loops reported in \citet{2021A&A...652L...3M} have not been provided but measured in our study.}
    \label{fig:scale}
\end{figure}

Fig.~\ref{fig:scale}(a) shows the scaling of the loop length and oscillation periods obtained in this study and previous reports of decayless kink oscillations of coronal loops \citep[e.g.,][]{2012ApJ...751L..27W,2013A&A...552A..57N,2013A&A...560A.107A, 2015A&A...583A.136A,2018ApJ...854L...5D,2019ApJ...884L..40A,2021A&A...652L...3M,2022MNRAS.513.1834Z,2022arXiv220505319P}. 
As mentioned in Section~\ref{sec:co}, the oscillation periods in our work are rather short ($<3$~min), so our data points lie in the lower left corner of the scaling plot. 
The previously established linear scaling of the loop length and period is not broken down by our results, as demonstrated by the black line, the best linear fit with the gradient of about 1.0\,s\,Mm$^{-1}$. 

In our observations, only certain segments near the loop top are clearly seen, see Fig.~\ref{fig:evo}. Thus we cannot identify the exact locations of the footpoints of oscillating threads, and take two polarities connected by this set of loops as approximate footpoints in the length estimation. The shaded box in Fig.~\ref{fig:scale}(a) indicates the uncertainty in the loop length.
We need to note that the period depends also on the local kink speed, which should explain the data scattering. During the observation, the oscillating loops are evolving, hence the instantaneous local kink speed is changing too. 
Also, a histogram of periods overplotted with the cumulative probability is shown in Fig.~\ref{fig:scale}(b). 

\section{Discussion}
\label{sec:discuss} 

Using HRI and AIA imaging data, we analyse decayless kink oscillations of several loops anchored at a quiescent active region.
In similar observational channels (171\,\AA\ of AIA and 174\,\AA\ of HRI), and with almost parallel LoS, the dynamics of loops are found to be visually identical in these two data sets. 

The observed oscillation signals are short-lasting, with amplitude below 0.1\,Mm and periods shorter than 2\,min.
The AIA signals delay for 66\,s compared to HRI because of different distances from the Sun. 
After the time correction, cross-correlation coefficients between oscillation signals in two data sets are typically higher than 0.9, suggesting oscillations recorded by two instruments are well consistent spatially and temporally.
Therefore, the observed oscillations are of the physical nature. 

\subsection{Detection of short time scales}
\label{sec:short} 

The period of analysed oscillations varies from 67\,s to 133\,s, which is rather short in comparison to previously reported cases. The uncertainties in the period estimation in Table~\ref{tab:pars} are connected with a relatively short duration of the oscillation detection (for some signals we only have a dozen of the loop displacement measurements), choice of the model to fit (the observed oscillatory signals may not be well described by a sine function), and the effects of slowly-varying background, surrounding threads, and noise.

In Fig.~\ref{fig:scale}(b), the number of cases with periods longer than 6\,min or shorter than 3\,min is 39, which is smaller than a third of the total number of 116 cases. On one hand, long periodicities (>6\,min) are possibly underreported
because the long loops ($>400$\,Mm) are rare and short-lived. Moreover, the detection of oscillations requires the preferable observational conditions, e.g., the contrasted boundary of the oscillating loop, to exist for longer than three oscillation period, which is less possible for longer periods. 
On the other hand, short periodicities
are rarely detected due to (1) the limited resolution (both temporal and spatial) of available instruments; (2) shorter loops are usually not well isolated and they evolve rapidly.

Since the period of kink oscillations scale linearly with the loop length (assuming for a certain parallel  harmonics), short periodicity can be formed in the shorter loop, that is, loop with lower height. Shorter plasma structures usually occur at the small-scale active regions (e.g., newly emerging ones) or short-distance magnetic polarity-pairs.
Therefore, short-period decayless kink oscillation possesses the potential for seismological diagnostics of the confined plasma at some certain coronal environments.
Recently, \citet{2022ApJ...930...55G} reported decayless kink oscillations of loops (averaged loop length is 23\,Mm) in mini-active regions, and successfully applied these oscillations to estimate Alfv\'en speed, magnetic field strength of the hosting loops.
In addition, short periodicities are of interest for the study of higher parallel harmonics of kink oscillations, and nonlinear cascade, e.g., the Kelvin-Helmholtz instability \citep[e.g.,][]{2008ApJ...687L.115T,2017SoPh..292..111R,2018ApJ...853...35T}.

The two-spacecraft detection of short periodicity in our work moves a step closer to unexplored regime of shorter oscillation periods, e.g., shorter than 1 minute. It shows that our current methods are applicable for short-period oscillations taken with EUI data. 
In particular, a candidate event of short-period decayless oscillations taken by EUI was observed by \citet{2022arXiv220505319P}, with oscillation periods of 14\,s and 30\,s in two short loops of 4.5\,Mm and 11\,Mm respectively. Moreover, in addition to the discussed kink oscillations, HRI with time cadence up to 2\,s could resolve such short period wave phenomena as sausage oscillations of flare loops with period of tens of seconds \citep[e.g.,][and references therein]{2020SSRv..216..136L}. 

\subsection{The problem of alignment}
\label{sec:aligndisc}

HRI images are affected by the spacecraft jitter. 
Some of this jitter (but not all) is measured by the spacecraft itself and recorded in the metadata of the EUI images (FITS World Coordinate System keywords). For the EUI Full Sun Imager (FSI), these pointing information is further updated on the ground with a limb fitting procedure. For the partial FOV images taken by HRI this is however not possible and only the spacecraft provided pointing information is available.
Therefore additional alignments internal to the analysed data cube
are needed to get rid of the remaining spacecraft jitter.
For the detection of oscillations of low amplitude and short period, the accuracy of the alignment is crucial as inaccurate alignment may lead to the appearance of artificial oscillations.

As mentioned in Section.~\ref{sec:align1}, the correlation-based internal alignment could achieve sub-pixel accuracy.
In practice, other factors affecting the accuracy include the selection of the correlated region (location and size), and the frames correlated at one time etc. 
In addition, it is found that the image enhancement prior to the alignment helps to improve accuracy \citep[e.g.,][]{1974Pratt,2021MNRAS.503.5715L}.
In the final step of alignment, 2D shifting with non-integer displacement by interpolation would degrade the spatial resolution of images. However, the loss of information is very minor. The high correlation of small-amplitude kink oscillations detected with HRI and AIA indicates that this internal alignment procedure is highly robust.

\section{Conclusions}
\label{sec:conclu}

In this study, short-period decayless kink oscillations of an ensemble of loops are investigated, utilizing EUI-HRI 174\,\AA\ image sequence and co-aligned AIA observation in the 171\,\AA\ channel. The analysis of imaging data is performed with the motion magnification technique.
The observed oscillating loops have average length of 51\,Mm, oscillation periods are about 60-100~s, and displacement amplitudes of 27--83\,km.
Given that EUI and AIA have different distances from the Sun, oscillations captured by AIA are delayed by 66\,s. After compensation of this time lag, oscillations obtained in two data are correlated well with each other, with their cross correlation coefficients ranging from 0.82 to 0.97.
It is confirmed that EUI-HRI sees the same oscillations as AIA, a well-tested instrument, which is a starting point for progressing to the study of shorter oscillation periods with HRI.

\section*{Acknowledgements}
Solar Orbiter is a mission of internal cooperation between ESA and NASA, operated by ESA. 
The EUI instrument was built by CSL, IAS, MPS, MSSL/UCL, PMOD/WRC, ROB, LCF/IO with funding from the Belgian Federal Science Policy Office (BELSPO/PRODEX PEA C4000134088); the Centre National d’Etudes Spatiales (CNES); the UK Space Agency (UKSA); the Bundesministerium für Wirtschaft und Energie (BMWi) through the Deutsches Zentrum für Luft- und Raumfahrt (DLR); and the Swiss Space Office (SSO).
SDO data supplied courtesy of the SDO/HMI and SDO/AIA consortia. 
S.Z. is supported by the China Scholarship Council--University of Warwick joint scholarship. V.M.N. and D.Y.K. acknowledge support
from the STFC Consolidated Grant ST/T000252/1.

\section*{Data Availability}


In this paper, we analysed data using the Interactive Data Language (IDL), SolarSoftWare (SSW, \citealt{1998SoPh..182..497F}) package, and the motion magnification technique \citep{2016SoPh..291.3251A}, and the Solar Bayesian Analysis Toolkit (SoBAT, \citealt{2021ApJS..252...11A}).
The EUI data are available at DOI: \url{https://doi.org/10.24414/2qfw-tr95}. The AIA and HMI data are available at \url{http://jsoc.stanford.edu/}.
The data are processed and analysed using the routines available at \url{https://www.lmsal.com/sdodocs/doc/dcur/SDOD0060.zip/zip/entry/} (Section 7).
The motion magnification code is available at \url{https://github.com/Sergey-Anfinogentov/motion_magnification}. 



\bibliographystyle{mnras}

\begin{thebibliography}{}
\makeatletter
\relax
\def\mn@urlcharsother{\let\do\@makeother \do\$\do\&\do\#\do\^\do\_\do\%\do\~}
\def\mn@doi{\begingroup\mn@urlcharsother \@ifnextchar [ {\mn@doi@}
  {\mn@doi@[]}}
\def\mn@doi@[#1]#2{\def\@tempa{#1}\ifx\@tempa\@empty \href
  {http://dx.doi.org/#2} {doi:#2}\else \href {http://dx.doi.org/#2} {#1}\fi
  \endgroup}
\def\mn@eprint#1#2{\mn@eprint@#1:#2::\@nil}
\def\mn@eprint@arXiv#1{\href {http://arxiv.org/abs/#1} {{\tt arXiv:#1}}}
\def\mn@eprint@dblp#1{\href {http://dblp.uni-trier.de/rec/bibtex/#1.xml}
  {dblp:#1}}
\def\mn@eprint@#1:#2:#3:#4\@nil{\def\@tempa {#1}\def\@tempb {#2}\def\@tempc
  {#3}\ifx \@tempc \@empty \let \@tempc \@tempb \let \@tempb \@tempa \fi \ifx
  \@tempb \@empty \def\@tempb {arXiv}\fi \@ifundefined
  {mn@eprint@\@tempb}{\@tempb:\@tempc}{\expandafter \expandafter \csname
  mn@eprint@\@tempb\endcsname \expandafter{\@tempc}}}

\bibitem[\protect\citeauthoryear{{Afanasyev}, {Van Doorsselaere}  \&
  {Nakariakov}}{{Afanasyev} et~al.}{2020}]{2020A&A...633L...8A}
{Afanasyev} A.~N.,  {Van Doorsselaere} T.,   {Nakariakov} V.~M.,  2020, \mn@doi
  [\aap] {10.1051/0004-6361/201937187}, \href
  {https://ui.adsabs.harvard.edu/abs/2020A&A...633L...8A} {633, L8}

\bibitem[\protect\citeauthoryear{Almonacid-Caballer, Pardo-Pascual  \&
  Ruiz}{Almonacid-Caballer et~al.}{2017}]{2017Jaime}
Almonacid-Caballer J.,  Pardo-Pascual J.~E.,   Ruiz L.~A.,  2017, \mn@doi
  [Remote Sensing] {10.3390/rs9101051}, 9

\bibitem[\protect\citeauthoryear{{Andries}, {Arregui}  \& {Goossens}}{{Andries}
  et~al.}{2005}]{2005ApJ...624L..57A}
{Andries} J.,  {Arregui} I.,   {Goossens} M.,  2005, \mn@doi [\apjl]
  {10.1086/430347}, \href
  {https://ui.adsabs.harvard.edu/abs/2005ApJ...624L..57A} {624, L57}

\bibitem[\protect\citeauthoryear{{Andries}, {van Doorsselaere}, {Roberts},
  {Verth}, {Verwichte}  \& {Erd{\'e}lyi}}{{Andries}
  et~al.}{2009}]{2009SSRv..149....3A}
{Andries} J.,  {van Doorsselaere} T.,  {Roberts} B.,  {Verth} G.,  {Verwichte}
  E.,   {Erd{\'e}lyi} R.,  2009, \mn@doi [\ssr] {10.1007/s11214-009-9561-2},
  \href {https://ui.adsabs.harvard.edu/abs/2009SSRv..149....3A} {149, 3}

\bibitem[\protect\citeauthoryear{{Anfinogentov} \& {Nakariakov}}{{Anfinogentov}
  \& {Nakariakov}}{2016}]{2016SoPh..291.3251A}
{Anfinogentov} S.,  {Nakariakov} V.~M.,  2016, \mn@doi [\solphys]
  {10.1007/s11207-016-1013-z}, \href
  {https://ui.adsabs.harvard.edu/abs/2016SoPh..291.3251A} {291, 3251}

\bibitem[\protect\citeauthoryear{{Anfinogentov} \& {Nakariakov}}{{Anfinogentov}
  \& {Nakariakov}}{2019}]{2019ApJ...884L..40A}
{Anfinogentov} S.~A.,  {Nakariakov} V.~M.,  2019, \mn@doi [\apjl]
  {10.3847/2041-8213/ab4792}, \href
  {https://ui.adsabs.harvard.edu/abs/2019ApJ...884L..40A} {884, L40}

\bibitem[\protect\citeauthoryear{{Anfinogentov}, {Nistic{\`o}}  \&
  {Nakariakov}}{{Anfinogentov} et~al.}{2013}]{2013A&A...560A.107A}
{Anfinogentov} S.,  {Nistic{\`o}} G.,   {Nakariakov} V.~M.,  2013, \mn@doi
  [\aap] {10.1051/0004-6361/201322094}, \href
  {https://ui.adsabs.harvard.edu/abs/2013A&A...560A.107A} {560, A107}

\bibitem[\protect\citeauthoryear{{Anfinogentov}, {Nakariakov}  \&
  {Nistic{\`o}}}{{Anfinogentov} et~al.}{2015}]{2015A&A...583A.136A}
{Anfinogentov} S.~A.,  {Nakariakov} V.~M.,   {Nistic{\`o}} G.,  2015, \mn@doi
  [\aap] {10.1051/0004-6361/201526195}, \href
  {https://ui.adsabs.harvard.edu/abs/2015A&A...583A.136A} {583, A136}

\bibitem[\protect\citeauthoryear{{Anfinogentov}, {Nakariakov}, {Pascoe}  \&
  {Goddard}}{{Anfinogentov} et~al.}{2021}]{2021ApJS..252...11A}
{Anfinogentov} S.~A.,  {Nakariakov} V.~M.,  {Pascoe} D.~J.,   {Goddard} C.~R.,
  2021, \mn@doi [\apjs] {10.3847/1538-4365/abc5c1}, \href
  {https://ui.adsabs.harvard.edu/abs/2021ApJS..252...11A} {252, 11}

\bibitem[\protect\citeauthoryear{{Antolin}, {De Moortel}, {Van Doorsselaere}
  \& {Yokoyama}}{{Antolin} et~al.}{2016}]{2016ApJ...830L..22A}
{Antolin} P.,  {De Moortel} I.,  {Van Doorsselaere} T.,   {Yokoyama} T.,  2016,
  \mn@doi [\apjl] {10.3847/2041-8205/830/2/L22}, \href
  {https://ui.adsabs.harvard.edu/abs/2016ApJ...830L..22A} {830, L22}

\bibitem[\protect\citeauthoryear{{Duckenfield}, {Anfinogentov}, {Pascoe}  \&
  {Nakariakov}}{{Duckenfield} et~al.}{2018}]{2018ApJ...854L...5D}
{Duckenfield} T.,  {Anfinogentov} S.~A.,  {Pascoe} D.~J.,   {Nakariakov} V.~M.,
   2018, \mn@doi [\apjl] {10.3847/2041-8213/aaaaeb}, \href
  {https://ui.adsabs.harvard.edu/abs/2018ApJ...854L...5D} {854, L5}

\bibitem[\protect\citeauthoryear{Feng, Deng, Shu, Wang, Deng  \& Ji}{Feng
  et~al.}{2012}]{6463241}
Feng S.,  Deng L.,  Shu G.,  Wang F.,  Deng H.,   Ji K.,  2012, in 2012 IEEE
  Fifth International Conference on Advanced Computational Intelligence
  (ICACI). pp 626--630, \mn@doi{10.1109/ICACI.2012.6463241}

\bibitem[\protect\citeauthoryear{{Freeland} \& {Handy}}{{Freeland} \&
  {Handy}}{1998}]{1998SoPh..182..497F}
{Freeland} S.~L.,  {Handy} B.~N.,  1998, \mn@doi [\solphys]
  {10.1023/A:1005038224881}, \href
  {https://ui.adsabs.harvard.edu/abs/1998SoPh..182..497F} {182, 497}

\bibitem[\protect\citeauthoryear{{Gao}, {Tian}, {Van Doorsselaere}  \&
  {Chen}}{{Gao} et~al.}{2022}]{2022ApJ...930...55G}
{Gao} Y.,  {Tian} H.,  {Van Doorsselaere} T.,   {Chen} Y.,  2022, \mn@doi
  [\apj] {10.3847/1538-4357/ac62cf}, \href
  {https://ui.adsabs.harvard.edu/abs/2022ApJ...930...55G} {930, 55}

\bibitem[\protect\citeauthoryear{Guizar-Sicairos, Thurman  \&
  Fienup}{Guizar-Sicairos et~al.}{2008}]{Guizar-Sicairos:2008}
Guizar-Sicairos M.,  Thurman S.~T.,   Fienup J.~R.,  2008, \mn@doi [Opt. Lett.]
  {10.1364/OL.33.000156}, 33, 156

\bibitem[\protect\citeauthoryear{{Guo}, {Van Doorsselaere}, {Karampelas}, {Li},
  {Antolin}  \& {De Moortel}}{{Guo} et~al.}{2019}]{2019ApJ...870...55G}
{Guo} M.,  {Van Doorsselaere} T.,  {Karampelas} K.,  {Li} B.,  {Antolin} P.,
  {De Moortel} I.,  2019, \mn@doi [\apj] {10.3847/1538-4357/aaf1d0}, \href
  {https://ui.adsabs.harvard.edu/abs/2019ApJ...870...55G} {870, 55}

\bibitem[\protect\citeauthoryear{{Karampelas} \& {Van
  Doorsselaere}}{{Karampelas} \& {Van
  Doorsselaere}}{2020}]{2020ApJ...897L..35K}
{Karampelas} K.,  {Van Doorsselaere} T.,  2020, \mn@doi [\apjl]
  {10.3847/2041-8213/ab9f38}, \href
  {https://ui.adsabs.harvard.edu/abs/2020ApJ...897L..35K} {897, L35}

\bibitem[\protect\citeauthoryear{{Karampelas} \& {Van
  Doorsselaere}}{{Karampelas} \& {Van
  Doorsselaere}}{2021}]{2021ApJ...908L...7K}
{Karampelas} K.,  {Van Doorsselaere} T.,  2021, \mn@doi [\apjl]
  {10.3847/2041-8213/abdc2b}, \href
  {https://ui.adsabs.harvard.edu/abs/2021ApJ...908L...7K} {908, L7}

\bibitem[\protect\citeauthoryear{{Karampelas}, {Van Doorsselaere}, {Pascoe},
  {Guo}  \& {Antolin}}{{Karampelas} et~al.}{2019}]{2019FrASS...6...38K}
{Karampelas} K.,  {Van Doorsselaere} T.,  {Pascoe} D.~J.,  {Guo} M.,
  {Antolin} P.,  2019, \mn@doi [Frontiers in Astronomy and Space Sciences]
  {10.3389/fspas.2019.00038}, \href
  {https://ui.adsabs.harvard.edu/abs/2019FrASS...6...38K} {6, 38}

\bibitem[\protect\citeauthoryear{Lehmann, Gonner  \& Spitzer}{Lehmann
  et~al.}{1999}]{Lehmann1999}
Lehmann T.,  Gonner C.,   Spitzer K.,  1999, \mn@doi [IEEE Transactions on
  Medical Imaging] {10.1109/42.816070}, 18, 1049

\bibitem[\protect\citeauthoryear{{Lemen} et~al.,}{{Lemen}
  et~al.}{2012}]{2012SoPh..275...17L}
{Lemen} J.~R.,  et~al., 2012, \mn@doi [\solphys] {10.1007/s11207-011-9776-8},
  \href {https://ui.adsabs.harvard.edu/abs/2012SoPh..275...17L} {275, 17}

\bibitem[\protect\citeauthoryear{{Li}, {Antolin}, {Guo}, {Kuznetsov}, {Pascoe},
  {Van Doorsselaere}  \& {Vasheghani Farahani}}{{Li}
  et~al.}{2020}]{2020SSRv..216..136L}
{Li} B.,  {Antolin} P.,  {Guo} M.~Z.,  {Kuznetsov} A.~A.,  {Pascoe} D.~J.,
  {Van Doorsselaere} T.,   {Vasheghani Farahani} S.,  2020, \mn@doi [\ssr]
  {10.1007/s11214-020-00761-z}, \href
  {https://ui.adsabs.harvard.edu/abs/2020SSRv..216..136L} {216, 136}

\bibitem[\protect\citeauthoryear{{Liang}, {Qu}, {Chen}, {Zhong}, {Song}  \&
  {Li}}{{Liang} et~al.}{2021}]{2021MNRAS.503.5715L}
{Liang} Y.,  {Qu} Z.~Q.,  {Chen} Y.~J.,  {Zhong} Y.,  {Song} Z.~M.,   {Li}
  S.~Y.,  2021, \mn@doi [\mnras] {10.1093/mnras/stab463}, \href
  {https://ui.adsabs.harvard.edu/abs/2021MNRAS.503.5715L} {503, 5715}

\bibitem[\protect\citeauthoryear{{Mandal}, {Tian}  \& {Peter}}{{Mandal}
  et~al.}{2021}]{2021A&A...652L...3M}
{Mandal} S.,  {Tian} H.,   {Peter} H.,  2021, \mn@doi [\aap]
  {10.1051/0004-6361/202141542}, \href
  {https://ui.adsabs.harvard.edu/abs/2021A&A...652L...3M} {652, L3}

\bibitem[\protect\citeauthoryear{{M{\"u}ller} et~al.,}{{M{\"u}ller}
  et~al.}{2020}]{2020A&A...642A...1M}
{M{\"u}ller} D.,  et~al., 2020, \mn@doi [\aap] {10.1051/0004-6361/202038467},
  \href {https://ui.adsabs.harvard.edu/abs/2020A&A...642A...1M} {642, A1}

\bibitem[\protect\citeauthoryear{{Nakariakov}, {Aschwanden}  \& {van
  Doorsselaere}}{{Nakariakov} et~al.}{2009}]{2009A&A...502..661N}
{Nakariakov} V.~M.,  {Aschwanden} M.~J.,   {van Doorsselaere} T.,  2009,
  \mn@doi [\aap] {10.1051/0004-6361/200810847}, \href
  {https://ui.adsabs.harvard.edu/abs/2009A&A...502..661N} {502, 661}

\bibitem[\protect\citeauthoryear{{Nakariakov}, {Anfinogentov}, {Nistic{\`o}}
  \& {Lee}}{{Nakariakov} et~al.}{2016}]{2016A&A...591L...5N}
{Nakariakov} V.~M.,  {Anfinogentov} S.~A.,  {Nistic{\`o}} G.,   {Lee} D.~H.,
  2016, \mn@doi [\aap] {10.1051/0004-6361/201628850}, \href
  {https://ui.adsabs.harvard.edu/abs/2016A&A...591L...5N} {591, L5}

\bibitem[\protect\citeauthoryear{{Nakariakov} et~al.,}{{Nakariakov}
  et~al.}{2021}]{2021SSRv..217...73N}
{Nakariakov} V.~M.,  et~al., 2021, \mn@doi [\ssr] {10.1007/s11214-021-00847-2},
  \href {https://ui.adsabs.harvard.edu/abs/2021SSRv..217...73N} {217, 73}

\bibitem[\protect\citeauthoryear{{Nistic{\`o}}, {Nakariakov}  \&
  {Verwichte}}{{Nistic{\`o}} et~al.}{2013}]{2013A&A...552A..57N}
{Nistic{\`o}} G.,  {Nakariakov} V.~M.,   {Verwichte} E.,  2013, \mn@doi [\aap]
  {10.1051/0004-6361/201220676}, \href
  {https://ui.adsabs.harvard.edu/abs/2013A&A...552A..57N} {552, A57}

\bibitem[\protect\citeauthoryear{{Pesnell}, {Thompson}  \&
  {Chamberlin}}{{Pesnell} et~al.}{2012}]{2012SoPh..275....3P}
{Pesnell} W.~D.,  {Thompson} B.~J.,   {Chamberlin} P.~C.,  2012, \mn@doi
  [\solphys] {10.1007/s11207-011-9841-3}, \href
  {https://ui.adsabs.harvard.edu/abs/2012SoPh..275....3P} {275, 3}

\bibitem[\protect\citeauthoryear{{Petrova}, {Magyar}, {Van Doorsselaere}  \&
  {Berghmans}}{{Petrova} et~al.}{2022}]{2022arXiv220505319P}
{Petrova} E.,  {Magyar} N.,  {Van Doorsselaere} T.,   {Berghmans} D.,  2022,
  arXiv e-prints, \href {https://ui.adsabs.harvard.edu/abs/2022arXiv220505319P}
  {p. arXiv:2205.05319}

\bibitem[\protect\citeauthoryear{Pratt}{Pratt}{1974}]{1974Pratt}
Pratt W.~K.,  1974, \mn@doi [IEEE Transactions on Aerospace and Electronic
  Systems] {10.1109/TAES.1974.307828}, AES-10, 353

\bibitem[\protect\citeauthoryear{{Rochus} et~al.,}{{Rochus}
  et~al.}{2020}]{2020A&A...642A...8R}
{Rochus} P.,  et~al., 2020, \mn@doi [\aap] {10.1051/0004-6361/201936663}, \href
  {https://ui.adsabs.harvard.edu/abs/2020A&A...642A...8R} {642, A8}

\bibitem[\protect\citeauthoryear{{Ruderman}}{{Ruderman}}{2017}]{2017SoPh..292..111R}
{Ruderman} M.~S.,  2017, \mn@doi [\solphys] {10.1007/s11207-017-1133-0}, \href
  {https://ui.adsabs.harvard.edu/abs/2017SoPh..292..111R} {292, 111}

\bibitem[\protect\citeauthoryear{{Ruderman} \& {Petrukhin}}{{Ruderman} \&
  {Petrukhin}}{2021}]{2021MNRAS.501.3017R}
{Ruderman} M.~S.,  {Petrukhin} N.~S.,  2021, \mn@doi [\mnras]
  {10.1093/mnras/staa3816}, \href
  {https://ui.adsabs.harvard.edu/abs/2021MNRAS.501.3017R} {501, 3017}

\bibitem[\protect\citeauthoryear{{Ruderman}, {Petrukhin}  \&
  {Pelinovsky}}{{Ruderman} et~al.}{2021}]{2021SoPh..296..124R}
{Ruderman} M.~S.,  {Petrukhin} N.~S.,   {Pelinovsky} E.,  2021, \mn@doi
  [\solphys] {10.1007/s11207-021-01867-5}, \href
  {https://ui.adsabs.harvard.edu/abs/2021SoPh..296..124R} {296, 124}

\bibitem[\protect\citeauthoryear{{Scherrer} et~al.,}{{Scherrer}
  et~al.}{2012}]{2012SoPh..275..207S}
{Scherrer} P.~H.,  et~al., 2012, \mn@doi [\solphys]
  {10.1007/s11207-011-9834-2}, \href
  {https://ui.adsabs.harvard.edu/abs/2012SoPh..275..207S} {275, 207}

\bibitem[\protect\citeauthoryear{{Shi}, {Van Doorsselaere}, {Antolin}  \&
  {Li}}{{Shi} et~al.}{2021}]{2021ApJ...922...60S}
{Shi} M.,  {Van Doorsselaere} T.,  {Antolin} P.,   {Li} B.,  2021, \mn@doi
  [\apj] {10.3847/1538-4357/ac2497}, \href
  {https://ui.adsabs.harvard.edu/abs/2021ApJ...922...60S} {922, 60}

\bibitem[\protect\citeauthoryear{{Terradas}, {Andries}, {Goossens}, {Arregui},
  {Oliver}  \& {Ballester}}{{Terradas} et~al.}{2008}]{2008ApJ...687L.115T}
{Terradas} J.,  {Andries} J.,  {Goossens} M.,  {Arregui} I.,  {Oliver} R.,
  {Ballester} J.~L.,  2008, \mn@doi [\apjl] {10.1086/593203}, \href
  {https://ui.adsabs.harvard.edu/abs/2008ApJ...687L.115T} {687, L115}

\bibitem[\protect\citeauthoryear{{Terradas}, {Magyar}  \& {Van
  Doorsselaere}}{{Terradas} et~al.}{2018}]{2018ApJ...853...35T}
{Terradas} J.,  {Magyar} N.,   {Van Doorsselaere} T.,  2018, \mn@doi [\apj]
  {10.3847/1538-4357/aa9d0f}, \href
  {https://ui.adsabs.harvard.edu/abs/2018ApJ...853...35T} {853, 35}

\bibitem[\protect\citeauthoryear{{Tian}, {McIntosh}, {Wang}, {Ofman}, {De
  Pontieu}, {Innes}  \& {Peter}}{{Tian} et~al.}{2012}]{2012ApJ...759..144T}
{Tian} H.,  {McIntosh} S.~W.,  {Wang} T.,  {Ofman} L.,  {De Pontieu} B.,
  {Innes} D.~E.,   {Peter} H.,  2012, \mn@doi [\apj]
  {10.1088/0004-637X/759/2/144}, \href
  {https://ui.adsabs.harvard.edu/abs/2012ApJ...759..144T} {759, 144}

\bibitem[\protect\citeauthoryear{{Van Doorsselaere} et~al.,}{{Van Doorsselaere}
  et~al.}{2020}]{2020SSRv..216..140V}
{Van Doorsselaere} T.,  et~al., 2020, \mn@doi [\ssr]
  {10.1007/s11214-020-00770-y}, \href
  {https://ui.adsabs.harvard.edu/abs/2020SSRv..216..140V} {216, 140}

\bibitem[\protect\citeauthoryear{{Wang}, {Ofman}, {Davila}  \& {Su}}{{Wang}
  et~al.}{2012}]{2012ApJ...751L..27W}
{Wang} T.,  {Ofman} L.,  {Davila} J.~M.,   {Su} Y.,  2012, \mn@doi [\apjl]
  {10.1088/2041-8205/751/2/L27}, \href
  {https://ui.adsabs.harvard.edu/abs/2012ApJ...751L..27W} {751, L27}

\bibitem[\protect\citeauthoryear{{Yang}, {Qu}, {Ji}, {Feng}, {Deng}, {Lin}  \&
  {Wang}}{{Yang} et~al.}{2015}]{2015RAA....15..569Y}
{Yang} Y.-F.,  {Qu} H.-X.,  {Ji} K.-F.,  {Feng} S.,  {Deng} H.,  {Lin} J.-B.,
  {Wang} F.,  2015, \mn@doi [Research in Astronomy and Astrophysics]
  {10.1088/1674-4527/15/4/009}, \href
  {https://ui.adsabs.harvard.edu/abs/2015RAA....15..569Y} {15, 569}

\bibitem[\protect\citeauthoryear{{Zhong}, {Duckenfield}, {Nakariakov}  \&
  {Anfinogentov}}{{Zhong} et~al.}{2021}]{2021SoPh..296..135Z}
{Zhong} S.,  {Duckenfield} T.~J.,  {Nakariakov} V.~M.,   {Anfinogentov} S.~A.,
  2021, \mn@doi [\solphys] {10.1007/s11207-021-01870-w}, \href
  {https://ui.adsabs.harvard.edu/abs/2021SoPh..296..135Z} {296, 135}

\bibitem[\protect\citeauthoryear{{Zhong}, {Nakariakov}, {Kolotkov}  \&
  {Anfinogentov}}{{Zhong} et~al.}{2022}]{2022MNRAS.513.1834Z}
{Zhong} S.,  {Nakariakov} V.~M.,  {Kolotkov} D.~Y.,   {Anfinogentov} S.~A.,
  2022, \mn@doi [\mnras] {10.1093/mnras/stac1014}, \href
  {https://ui.adsabs.harvard.edu/abs/2022MNRAS.513.1834Z} {513, 1834}

\bibitem[\protect\citeauthoryear{Zitová \& Flusser}{Zitová \&
  Flusser}{2003}]{ZITOVA2003977}
Zitová B.,  Flusser J.,  2003, \mn@doi [Image and Vision Computing]
  {https://doi.org/10.1016/S0262-8856(03)00137-9}, 21, 977

\makeatother
\end{thebibliography}




\appendix

\section{Curve Fitting}
\label{appendix}
\begin{figure}
	\includegraphics[width=\columnwidth]{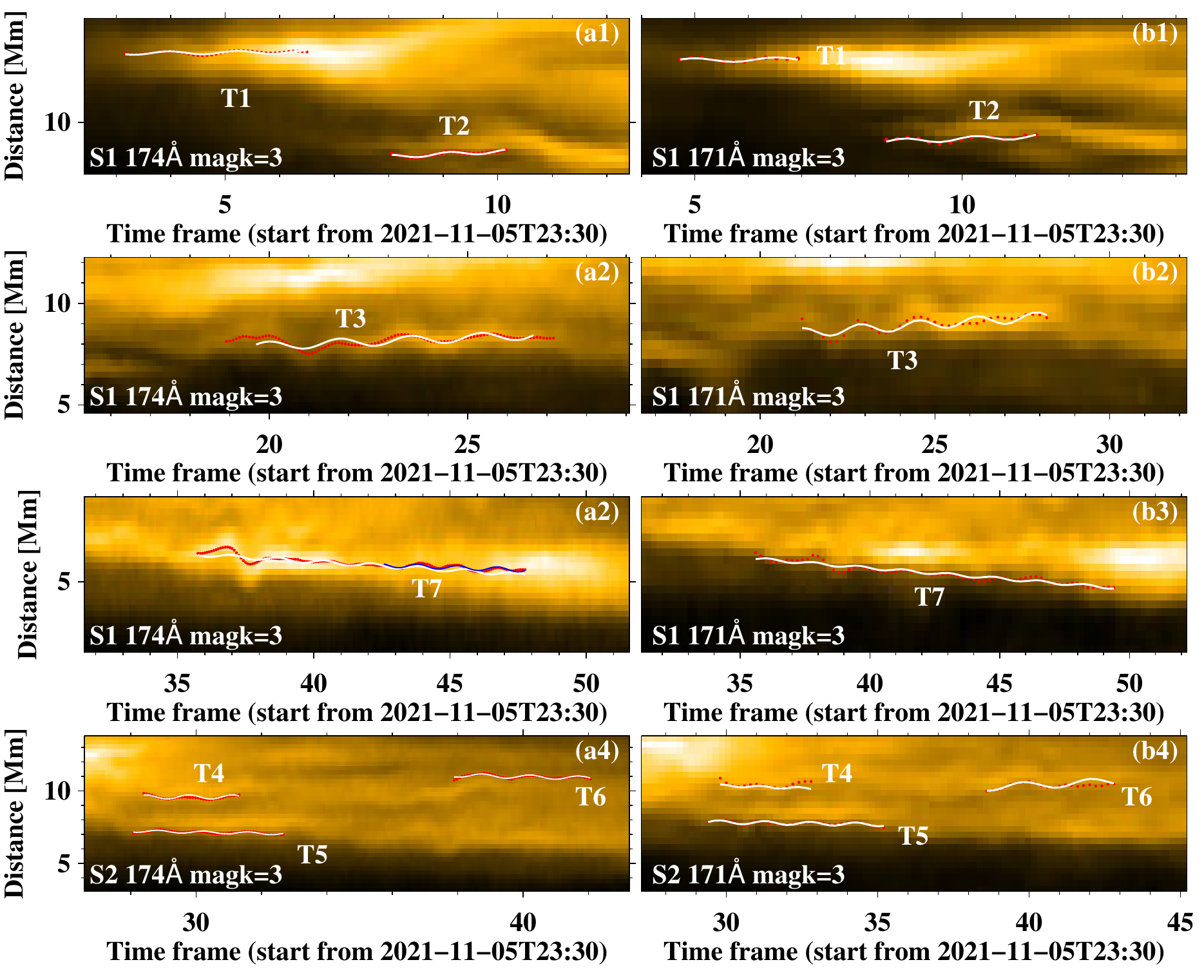}
    \caption{The analysed oscillation signals (red dots) obtained by HRI ((a1)-(a2)) and AIA ((b1)-(b4)) and their corresponding best sine fits (solid curves).}
    \label{fig:sinfit}
\end{figure}

Here we demonstrate the best sinusoidal fits of the oscillation signals of T1--T7 obtained in the HRI and AIA data. 
The signals (see the red dots in Fig.~\ref{fig:sinfit}) are best fitted with Eq.~\ref{eq:sine} to estimate the oscillation period and displacement amplitude, using the Markov Chain Monte Carlo method (\citealt{2021ApJS..252...11A}; the algorithm is available at \url{https://github.com/Sergey-Anfinogentov/SoBAT}).
The best fitting parameters are shown in Table~\ref{tab:pars} and the best fitting curves are represented by the solid curves in Fig.~\ref{fig:sinfit}.

\bsp	
\label{lastpage}
\end{document}